\UseRawInputEncoding
\documentclass[aps,prl,reprint,groupedaddress,amsmath,amssymb]{revtex4-1}
\usepackage{graphicx}
\usepackage{xcolor}
\usepackage{adjustbox}
\usepackage{dcolumn}
\usepackage{bm,comment}

\begin{document}

\title{Eddy diffusivity operator in homogeneous isotropic turbulence}
\author{Yasaman Shirian}
\email{yshirian@stanford.edu}

\author{Ali Mani}
\email{alimani@stanford.edu}
\affiliation{Center for Turbulence Research, Stanford University, Stanford, California 94305, USA}


\begin{abstract}
 We use the recently developed Macroscopic Forcing Method [Mani and Park, Physical Review Fluids, 6:054607, 2021] to compute the scale-dependent eddy diffusivity characterizing ensemble-averaged scalar and momentum transport in incompressible homogeneous isotropic turbulence. For scales larger than the energy containing eddies, eddy diffusivity is found to be constant and consistent with the Boussinesq approximation. However for small scales eddy diffusivity is found to vanish inversely proportional to the wavenumber.  Behavior at all scales is reasonably captured by a non-local eddy diffusivity operator modeled as $D/\sqrt{\mathcal{I}-l^2\nabla^2}$, where $D$ is the eddy diffusivity in the Boussinesq limit, and $l$ is a constant on the order of the large-eddy length.
    These results  present the first direct measurement of eddy diffusivity in turbulence with implications in turbulence modeling. 
\end{abstract}


\maketitle

The aim of a turbulence closure model is to accurately prescribe the Reynolds stress tensor representing the mean momentum flux due to the underlying turbulent fluctuations.  Most models follow the Boussinesq prescription in which Reynolds stress is assumed to be proportional to the local mean strain-rate, the constant of proportionality $\nu_t$ being the eddy viscosity.  In more sophisticated models, known as algebraic stress models \cite{Rodi} the stress is written as a tensorially consistent polynomial of the strain and rotation tensors.  In such models, the eddy viscosity can be function of space and time and given either algebraically or via additional transport equations (see the textbook by Pope \cite{Pope}).


In \cite{Park}, we developed a technique, called the Macroscopic Forcing Method (MFM) which reveals the form of turbulence closure operators from  analyses of appropriately forced direct simulations of turbulence.  This approach should be contrasted with the technique of \textit{a priori} testing in which a model form is assumed beforehand and its predictions for the turbulent stress compared statistically with the actual stresses in a direct simulation.


This report presents the first application of MFM for measurement of turbulence eddy viscosity operator for momentum transport, and eddy diffusivity operator for scalar transport in homogeneous isotropic turbulence (HIT). From here on we refer to both differential operators as ``eddy diffusivity," assuming for momentum transport it refers to eddy viscosity normalized by the fluid density.

We clarify that the eddy diffusivity operators presented here are not claimed to be universal models for turbulence closure. Specifically, in scenarios with significant statistical inhomogeneity and anisotropy, such as in wall bounded flows, the form of the eddy diffusivity operator is expected to be different from those measured here for HIT. Nevertheless, studying HIT has contributed valuable insights towards understanding of mixing by turbulence. As such, accurate quantitative determination of eddy diffusivity by this flow is a starting stepping stone towards understanding of model constraints and expected model forms for turbulence closure in broader settings. Most commonly analogies to HIT have been made for the case of free shear flows, and away from thin zones, where flow structures are fully developed, as well as for understanding of mixing by small scales.  To follow the same spirit, we present a followup analysis in which the quantified eddy diffusivity operator is adopted for Reynolds Averaged Navier-Stokes (RANS) analysis of a self similar turbulent round jet demonstrating substantial improvement in prediction of the mean velocity profile.  
We begin by providing a brief overview of MFM, which is discussed in depth by\cite{Park}, but tailored here specifically for applications to statistically stationary flows.

As a starting point, it is more intuitive to discuss the methodology first in the context of scalar transport, by considering the advection diffusion equation,
\begin{equation}
\label{eq:scalar}
    \frac{\partial c}{\partial t}+\frac{\partial}{\partial x_j}\left(u_jc\right)-D_M\frac{\partial^2c}{\partial x_j \partial x_j}=0,
\end{equation}
where $u_j$ is the given turbulent flow field, $c$ is the scalar field, and $D_M$ is the molecular diffusivity. Equation (\ref{eq:scalar}) is linear and can be compactly expressed as $\mathcal{L}c=0$, where $\mathcal{L}$ is the advection-diffusion operator. We next consider the special case of statistically time stationary flows, and define the ensemble averaged concentration, denoted by $\overline{c}$, as time average of $c$ over a long time. Mathematically this is a linear projection denoted as $\overline{c}=Pc$. 

The task of turbulence modeling is to provide a low-dimensional operator that can directly predict $\overline{c}$ without the need to solve (\ref{eq:scalar}). We denote the sets of all possible solutions for $c$ and $\overline{c}$ as microscopic and macroscopic space respectively, noting the latter is significantly lower dimensional than the former. The advection-diffusion operator, $\mathcal{L}$, is the microscopic operator with its domain and range both in the microscopic space. The macroscopic operator denoting the turbulence model, which we call $\overline{\mathcal{L}}$, is a scaled-up operator defined in the macroscopic space such that
\begin{equation}
\label{eq:macroscopic}
    \overline{\mathcal{L}}\overline{c}=0.
\end{equation}

One can show that  $\overline{\mathcal{L}}$ can be directly determined from $\mathcal{L}$ as $\overline{\mathcal{L}}=\left(P\mathcal{L}^{-1}E\right)^{-1}$,\cite{Park} where $E$ is the linear extension operator that interpolates macroscopic fields back into the microscopic space. Direct determination of $\overline{\mathcal{L}}$ from this approach, however, requires inversion of the microscopic operator, which is often computationally infeasible.


MFM is a remedy, that allows determination of $\overline{\mathcal{L}}$ without inversion of $\mathcal{L}$. Specifically, this method applies forcing, $s$, to Equation (\ref{eq:macroscopic}),
\begin{equation}
\label{eq:MFM}
    \overline{\mathcal{L}}\overline{c}=s.
\end{equation}
and determines $\overline{\mathcal{L}}$ via examination of its response to various forcing scenarios. Given Equations (\ref{eq:macroscopic}) and (\ref{eq:MFM}) are written in the macroscopic space, the applied forcing must be macroscopic. For statistically stationary flows a time-independent $s$ would be appropriate for determination of the steady form of the closure operators.

 In practice, for any force field $s(x,y,z)$ in (\ref{eq:MFM}),  one can examine the direct solution to (\ref{eq:scalar}) subject to the same force field, 
\begin{equation}
\label{eq:scalarsource}
    \frac{\partial c}{\partial t}+\frac{\partial}{\partial x_j}\left(u_jc\right)-D_M\frac{\partial^2c}{\partial x_j \partial x_j}=s,
\end{equation}
and obtain $\overline{c}$ by projecting the response of (\ref{eq:scalarsource}) back to the macroscopic space. An input-output analysis extended over the entire macroscopic space, can precisely quantify $\overline{\mathcal{L}}$. 

\begin{figure}
    \centering
    \includegraphics[width=0.48\textwidth]{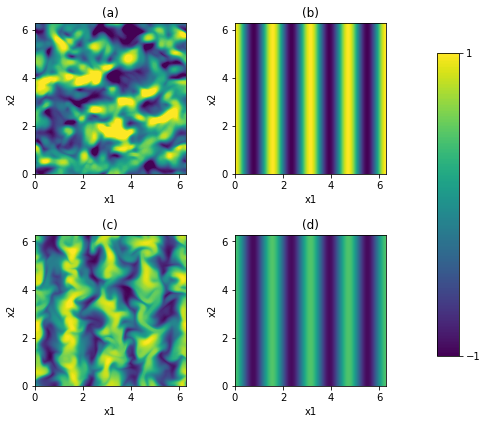}
    \caption{(a) A snapshot of the HIT simulation representing a component of the velocity field (b) The real part of macroscopic force, $s$, for mode $k=4$ (c) A 2D snapshot of the instantaneous scalar field, $c$, subject to the given flow and $s$ (d) The mean scalar field $\overline{c}$.}
    \label{fig:figure1}
\end{figure}
Given the statistical homogeneity of HIT, the macroscopic operator, $\overline{\mathcal{L}}$, must be a convolution in three spatial dimensions. It is therefore convenient to analyze (\ref{eq:MFM}) in Fourier space, by examining fields of $\overline{c}$ obtained from solutions to (\ref{eq:scalarsource}) subject to spatially harmonic source terms. Given the isotropy of the problem, it is sufficient to examine wavenumbers only in one direction, thus considering $s=\exp(ikx_1)$. It should be clarified that while imaginary concentration fields are unphysical, we here adopt such fields to ease the mathematical process of determining the eddy diffusivity operator. As we shall see, the final operator is expressible in the real space admitting physically realizable solutions.

We performed direct numerical simulation (DNS) of (\ref{eq:scalarsource}) for a wide range of $k$ values and measured the resulting $\overline{c}$ as shown in Figure~\ref{fig:figure1}. Given linearity and statistical homogeneity of the system, the resulting mean scalar field is expressible as $\overline{c}=\widehat{\overline{c}}\exp(ikx_1)$. In practice, due to finite time sampling, a small statistical noise exists in all 3D wavenumbers. One can then estimate $\widehat{\overline{c}}$ via projection of the numerical $\overline{c}$ to Fourier space and examining only the wavenumber ${\bf k}=(k,0,0)$. Taking Fourier transform of (\ref{eq:MFM}), the macroscopic operator can then be quantified as $\widehat{\overline{\mathcal{L}}}(k)=1/\widehat{\overline{c}}(k)$.

Having $\widehat{\overline{\mathcal{L}}}$ at hand, one can reveal an operator for the turbulence closure term as follows. Ensemble averaging of (\ref{eq:scalarsource}), provides the Reynolds-averaged transport equation as $\nabla \overline{{\bf u}^\prime c^\prime} - D_M \nabla^2\overline{c}=\overline{s}=s$. Here the prime ($^\prime$) superscript denotes mean-subtracted fields, and we note that $\overline{\bf u}=0$ in HIT. The first term on the left hand side of this equation, is the turbulence closure term that needs to be modeled via a closure operator as  $\nabla \overline{{\bf u}^\prime c^\prime}=\overline{\mathcal{L}^\prime}\overline{c}$. With this definition and given the ensemble-averaged equation,  $\overline{\mathcal{L}^\prime}$ can be expressed in terms of the measured $\overline{\mathcal{L}}$ as 
$\overline{\mathcal{L}^\prime}=\overline{\mathcal{L}}+D_M\nabla^2$. Expressing this relation in Fourier space and in terms of the measured $\widehat{\overline{c}}$ from the DNS solutions, one can write the turbulence closure operator as 
\begin{equation}
\label{eq:closure}
    \widehat{\overline{\mathcal{L}^\prime}}(k)=1/\widehat{\overline{c}}-D_Mk^2.
\end{equation}

The presented methodology above, can be extended to analyze momentum transport.  However, Navier-Stokes equation is not the appropriate framework for input-output analysis in this context. Similar to the case of scalar transport, in this case we address the question of how a ``{\it given}" flow transports momentum in the mean sense. The more foundational framework to address this question is therefore the Reynolds Transport Theorem  (RTT)\cite{Leal2007}, instead of Navier-Stokes. In a related context, various researchers have studied transport of passive vector fields by pre-specified velocity fields subject to the incompressibility constraint \cite{Arad2001,Angheluta2006,Benzi2001,Antonov2003}. The passive vector field equation can be derived from RTT by relaxing the constraint on the advective velocity, allowing it to be different from the transported momentum field itself. These prior studies, however, were not focused on determining eddy diffusivity operators. Doing so requires augmentation of equations with deterministic macroscopic forcing and examination of mean outcomes for various forcing scenarios as discussed for the case of scalar transport. To this end, Mani and Park\cite{Park} utilized the Generalized Momentum Transport (GMT) equation written as 
\begin{equation}
\label{eq:GMT}
    \frac{\partial v_i}{\partial t}+\frac{\partial}{\partial x_j}\left(u_jv_i\right)=-\frac{1}{\rho}\frac{\partial q}{\partial x_i}+\nu\frac{\partial^2v_i}{\partial x_j \partial x_j}+ s_i,
\end{equation}
where $u_i$ is a given velocity field satisfying the original Navier-Stokes equations. This velocity field, called the donor field, represents the flow whose eddy diffusivity is to be measured. $v_i$ represents momentum per unit mass transported by $u_i$. $\rho$ and $\nu$ are fluid density and molecular diffusivity, and $q$ is the pressure required to satisfy the constraint $\partial v_i/\partial x_i=0$. Equation~(\ref{eq:GMT}) is linear and therefore permits application of the same methodology as for passive scalar transport. Such a procedure will allow us to quantitatively determine how the given flow field ${\bf u}$ transports momentum in the mean sense. 


Similar to the case of scalar transport, the turbulent closure term,  $(\partial/\partial x_j)(u^\prime_j v^\prime_i)$, is desired to be expressed in terms of an operator acting on the mean momentum field $\overline{\bf v}$. In~\cite{Park} a proof is provided explaining why the turbulence closure operator for GMT is also a closure operator for the Navier-Stokes equation. Specifically, GMT is a robust generalization of the Navier-Stokes equation in the sense that under arbitrary initial conditions for $v_i$ the Navier-stokes solution, $u_i$, is GMT's stable attractor. This is true as long as GMT has consistent boundary conditions and its body force is dropped, since standard Navier-Stokes does not have a body force. 

Lastly, a quantified turbulent closure operator, $\overline{\mathcal{L}^\prime}$, can be related to the corresponding eddy diffusivity operator, $\mathcal{D}$, as $\overline{\mathcal{L}^\prime}=-\nabla \mathcal{D} \nabla$, given its definition. Expressing this result in Fourier space, allows quantification of the eddy diffusivity operator as $\widehat{\mathcal{D}}(k)=\widehat{\overline{\mathcal{L}^\prime}}(k)/k^2$. 

In order to avoid confusion between results of scalar and momentum transport analyses, we use the subscript $c$ to refer to operators associated with scalar transport and subscript $v$ to refer to operators associated with momentum transport. We next explain the details of the numerical procedure followed by the presentation of results.



\begin{table*}
\caption{\label{tab:input_parameters}%
Simulation parameters and flow statistics, reported uncertainties are 95\% statistical confidence level of the mean}
\begin{ruledtabular}
\begin{tabular}{ccccccc}
$Re_{\lambda}$ &  A & $\nu$ & $u_\text{rms}$ & $\varepsilon$ & $l_\text{eddy}$ & $T_\text{final}$/$t_\text{eddy}$\\
\hline
26 &  0.2792 &  0.0263 & 0.96$\pm$0.01 & 0.780$\pm$0.02 &1.15$\pm$0.06 &O(2000)\\
\hline
40 &  0.2792 &  0.0111 &  0.90$\pm$0.02 & 0.687$\pm$0.04 & 1.08$\pm$0.1  & O(500)\\
\hline
67 &  0.2792 & 0.0039 &  0.96$\pm$0.02 & 0.77$\pm$0.08 & 1.15$\pm$0.2 & O(100)\\
\end{tabular}
\end{ruledtabular}
\end{table*}
To generate the donor velocity fields, we performed DNS of incompressible homogeneous and isotropic turbulence in a triply periodic cubic domain of size $2\pi\times 2\pi\times 2\pi$ using uniform structured meshes. To sustain turbulence in a time-stationary fashion, we solved the Navier-Stokes equation subject to a forcing described by 
\begin{equation}
    \frac{\partial u_i}{\partial t}+\frac{\partial}{\partial x_j}\left(u_ju_i\right)=-\frac{1}{\rho}\frac{\partial p}{\partial x_i}+\nu\frac{\partial^2u_i}{\partial x_j \partial x_j}+ Au_i,
\end{equation}
where $A$ is the forcing constant. Table~\ref{tab:input_parameters} represents the nominal Reynolds numbers, $\text{Re}_\lambda$ for each simulation, and the specified $A$ and $\nu$ following the  prescription of \cite{Meneveau}. Additionally, the table lists the single-component velocity fluctuations $u_\text{rms}$, turbulent dissipation rate $\varepsilon=-\nu\overline{ u_i\nabla^2u_i}$, and the eddy size defined as $l_\text{eddy}=u_\text{rms}^3/\varepsilon$. The simulation times are reported in the table in units of $t_\text{eddy}=l_\text{eddy}/u_\text{rms}$. 

For these simulations, we adopted the code of \cite{Hadi} under the incompressible mode. The same code was used for numerical solutions to Equation~(\ref{eq:GMT}) after slight modification to allow $u_j$ and $v_i$ to be different vector fields and adjusting the forcing functions.  Additionally, for each flow field we solved (\ref{eq:scalarsource}) in order to analyze the closure operators for scalar fields. In this report we limit our studies to cases where $D_M=\nu$.  Given our focus on the fully developed regimes, the choice of the initial condition for  equations (\ref{eq:GMT}) and (\ref{eq:scalarsource}) is irrelevant. As such we simply used ${\bf v=0}$ and $c=0$ for these equations, respectively. In our post processing step the data associated with the initial transition of the transported fields are discarded. The time length of the discarded transitional regime was between 15 to 40 eddy turnover times, which was decided by monitoring the mean square of the transported quantities as a function of time.

The computational grids use $64^3$, $128^3$, and $256^3$ mesh points respectively for cases with $\text{Re}_\lambda=$ 26, 40 and 67. 
For each flow Reynolds number, equations  (\ref{eq:scalarsource}) and (\ref{eq:GMT}) are solved for $k$=0.25, 0.5, 1, 2, 4, and 8. To accommodate all wavenumbers, equations  (\ref{eq:scalarsource}) and (\ref{eq:GMT}) are solved on a computational domain with length $8\pi$ in the $x_1$ direction, where the velocity field  ${\bf u}$ is obtained by  the periodic extension of the nominal HIT solution for $x_1>2\pi$. 
\begin{figure*}[t]
    \centering
    \includegraphics[width=0.7\textwidth]{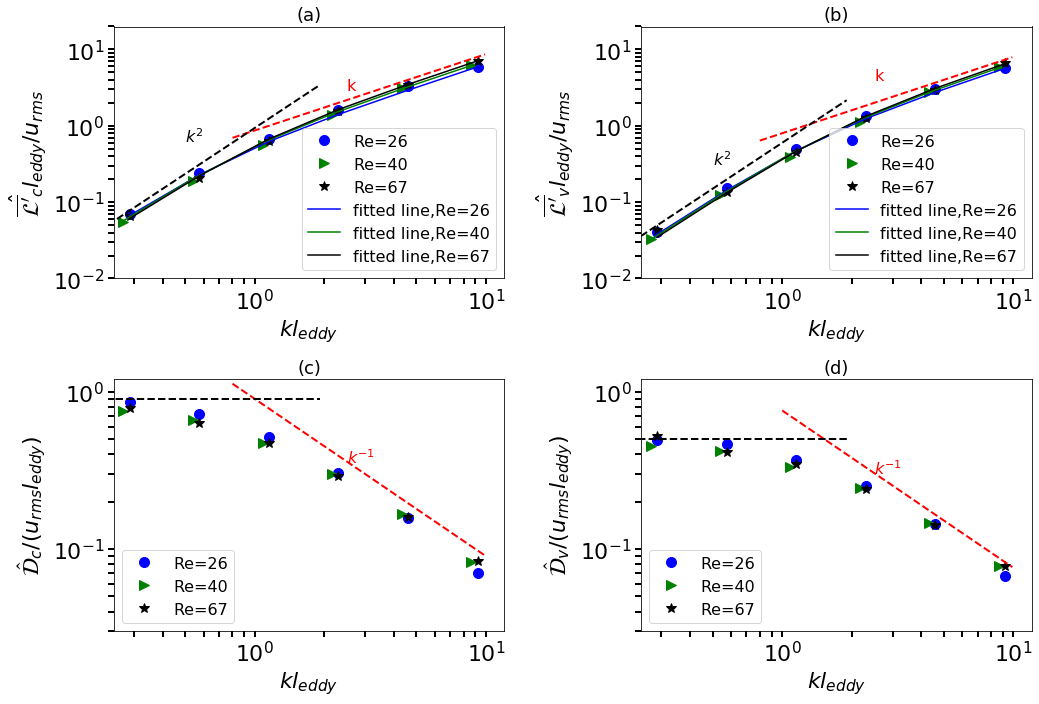}
    \caption{Measured macroscopic closure operators versus wavenumber for transport of (a) scalars and (b) momentum as well as the corresponding eddy diffusivity operators as shown in (c) and (d).}
    \label{fig:figure2}
\end{figure*}
In addition to these calculations, we analyzed cases to study the limit of $k=0$.  In this limit, the length scale of the mean field is infinite compared to the scale of the underlying turbulent flow structures. Therefore, the mixing process sees a locally linear mean profiles that can be approximated as $\overline {c}= a x_1 +b$. One can analytically show that inclusion of $b$ does not affect the fluctuating quantities. Furthermore, given the linearity of the transport equations it is sufficient to solely consider $a=1$ and thus $\overline{c}=x_1$. However, capturing this mean profile is not possible in periodic computational domains. Similar to what has been done in simulations of other periodic flows, e.g., inclusion of mean pressure gradient in turbulent channel flows \cite{kim1987}, one can analytically decompose the concentration field into a mean and fluctuating part as $c=x_1+c^\prime$. Substituting this decomposition back into the transport PDE, reveals a new PDE for $c'$ that is statistically homogeneous in all spatial directions, which can be solved on a periodic computational domain. In other words, MFM analysis of scalar transport in the limit of $k\rightarrow0$ collapses to methods described in homogenization theories \cite{Majda}. In this case a finite eddy diffusivity can be measured by quantification of the ratio of the mean turbulent flux and the imposed mean field gradient.

Figure~\ref{fig:figure2} shows the computed macroscopic closure operators as well as the corresponding eddy diffusivity operators for both scalar and momentum transport over a range of wavenumbers and Reynolds numbers. Scaling of data in units of $u_\text{rms}$ and $l_\text{eddy}$ results in reasonable collapse of data suggesting weak sensitivity to the Reynolds numbers. 
\begin{figure}[htb!]
    \centerline{
    \includegraphics[width=0.8\textwidth]{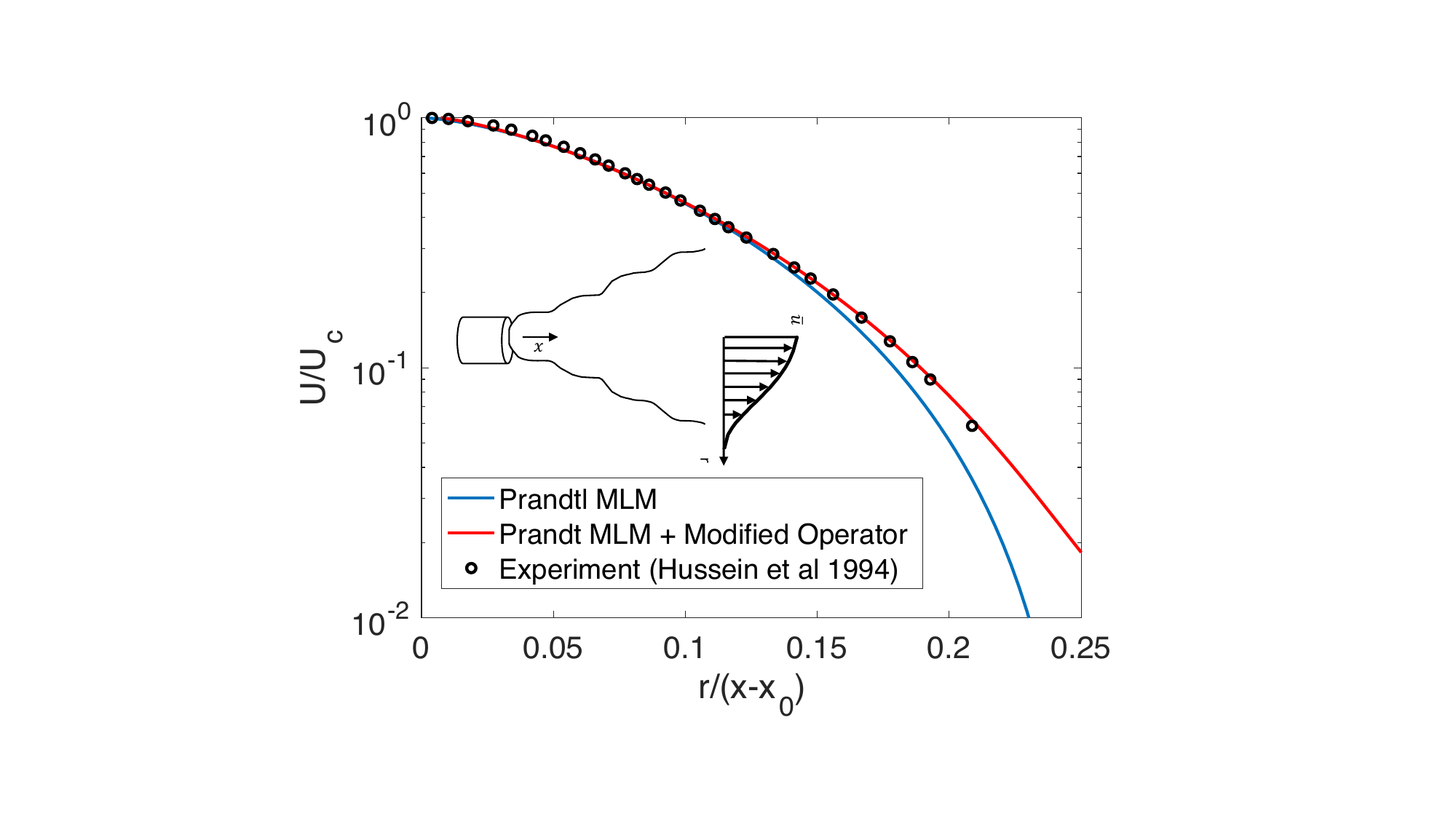}}
    \caption{Schematic of the self similar turbulent round jet and its mean velocity profile. Symbols show the Laser Doppler Anemometry (LDA) data of Hussein et al.\cite{Hussein1994}, the blue line shows RANS prediction using Prandtl mixing length model (PMLM), and the red line shows RANS prediction using a closure operator of the form shown in Equation (\ref{eq:D}) with coefficients scaled consistently with the PMLM prescription.}
    \label{fig:figure3_jet}
\end{figure}
All results in the limit of small $k$ indicate $\widehat{\overline{\mathcal{L}^\prime}}\sim k^2$, and a constant eddy diffusivity $\mathcal{D}\sim k^0$ consistent with the Boussinesq approximation \cite{Boussinesq}. In the large wavenumber limit, however, the closure operator significantly departs from that of Boussinesq limit. In this case, the eddy diffusivity drops inversely proportional to the wavenumber. This departure can be explained intuitively as follows: Standard diffusion implies that transported quantities propagate in space proportional to square root of time, $x\sim\sqrt{t}$, resulting in unbounded characteristic speed over short distance. However, given that here the underlying mechanism of transport is advection, transported quantities cannot propagate faster than linearly with time.  In fact, $k^1$ is an upper bound for a power law scaling of $\widehat{\overline{\mathcal{L}^\prime}}$ in the large $k$ limit as long as $k$ is below the limit where molecular diffusion would dominate advection in mixing. $\widehat{\overline{\mathcal{L}^\prime}}$ obtained from this study simply satisfies this physical limitation. 


\begin{table}[t]
\caption{\label{tab:fitted_coeffficients}%
Coefficients in the fitted curve ${D}k^2/{\sqrt{1+{(lk)}^2}}$ for macroscopic closure operator $\hat{\overline{{\mathcal{L}}^\prime}}\left(k\right)$. $D$ is reported in units of $u_\text{rms}l_\text{eddy}$, and $l$ is reported in units of $l_\text{eddy}$}
\begin{ruledtabular}
\begin{tabular}{ccccc}
                    & $D_c$ & $D_v$ & $l_c$ & $l_v$\\
\hline
$Re_{\lambda} = 26$ &  0.86 & 0.47 & 1.35 & 0.76\\
\hline
$Re_{\lambda} = 40$ &  0.82 & 0.45 & 1.14 & 0.66\\
\hline
$Re_{\lambda} = 67$ &  0.77 & 0.43 & 0.99 & 0.59\\
\end{tabular}
\end{ruledtabular}
\end{table}

Matching the two asymptotic limits of small and large $k$, we identify the following  expression as a uniformly valid approximation for the macroscopic closure operator 
\begin{equation}
 \widehat{\overline{\mathcal{L}^\prime}}=\frac{Dk^2}{\sqrt{1+l^2k^2}},   
\end{equation}
where the constants $D$ and $l$ are reported in Table~\ref{tab:fitted_coeffficients} and fitted to match the computed $\widehat{\overline{\mathcal{L}^\prime}}$ for $k \rightarrow 0$ and $k=8$. As shown in Figure~\ref{fig:figure2}a and b, this expression matches the numerically computed $\widehat{\overline{\mathcal{L}^\prime}}$ over all ranges of $k$. This operator can be expressed in physical space by replacing $k^2$ with $-\nabla^2$. The resulting operator is  $\overline{\mathcal{L}^\prime}=-\nabla \cdot \mathcal{D}\nabla$, where the eddy diffusivity operator, $\mathcal{D}$ can be expressed as
\begin{equation}
\label{eq:D}
    \mathcal{D}=\frac{D}{\sqrt{\mathcal{I}-l^2\nabla^2}},
\end{equation}
where $\mathcal{I}$ represents the identity operator. The denominator in (\ref{eq:D}) is an inverse operator, which indicates the non-locality of eddy diffusivity. $l$, which is on the order of the large eddy size, quantifies the extent of the non-locality. In other words, $l$ indicates how far the mean gradients at one location can influence mean fluxes at another location. This constant, plays a role similar to the mean free path in molecular systems, and following Prandtl, we refer to it as the mixing length.

The fact that $l_v$ is smaller than $l_c$ (see Table~\ref{tab:fitted_coeffficients}) indicates that the macroscopic closure operator is more local for momentum transport than for scalar transport. This is an unintuitive result, since at the DNS level, scalar transport is completely local, while momentum transport involves a non-local pressure projection. We conclude that at the macroscopic level, the non-locality of pressure cancels a portion of non-locality of advection. 

We next show the predictive impact of the obtained RANS operator by applying it to a practical flow and comparing its prediction against available experimental data. For this test case, we consider the self-similar turbulent round jet at a high Reynolds number. We then select an available eddy-diffusivity-based turbulence model, and replace its closure operator with Equation (\ref{eq:D}). For this purpose, we adopt the Prandtl Mixing Length Model (PMLM) \cite{Prandtl1925} since it provides $D$ and $l$ as explicit functions of spatial coordinates. We obtain solutions to both the original and modified RANS models with the sole difference that the eddy diffusivity in PMLM is replaced by the operator in Equation (\ref{eq:D}) using the same scaling for the coefficients as in PMLM. Details in implementation of both models are comprehensively described in the supplementary material \cite{supplemental}. Figure \ref{fig:figure3_jet} shows a remarkable improvement in RANS predictions, almost coinciding with the experimental data \cite{Hussein1994}.

Unlike most recent models, which rely on more complex sets of equations to provide the eddy diffusivity field, PMLM estimates the eddy diffusivity solely based on scaling analysis. The results presented in Figure \ref{fig:figure3_jet} highlight that significant improvement in model prediction can be achieved by merely improving the eddy diffusivity operator. 

It should be noted that the measured eddy diffusivity operator form based on HIT is unlikely to be universal. Specifically, wall bounded turbulent flows are known to involve significant anisotropy and inhomogeneity. The former effect results in a tensorial eddy viscosity and the latter results in a left-right asymmetry of non-local dependence of Reynolds stresses on the mean gradients in any given direction. Such non-localities cannot be captured by addition of Laplacian operators either directly or inversely in the eddy difusivity operator, since Laplacians inherently lead to symmetric Greens' functions. Both of these effects are missing in the closure operator obtained in this study. Likewise, the measured operator is unlikely to hold for multi-physics cases where the momentum equation is actively coupled with other transported quantities via body forces such as in natural convection. Even under linear coupling, in such cases macroscopic momentum transport is likely to be driven by mean gradients in both momentum and other actively coupled fields. 
The round jet problem, as a posteriori test,  was specifically chosen given it is a non wall bounded flow and uncoupled to any body force. 

 Perhaps the most remarkable outcome of the presented directly measured eddy diffusivity is its contrast to the scale-dependent eddy diffusivity predicted in theories based on renormalization group (RNG) \cite{Yakhot1986,Yakhot1992}. In these theories, assuming a Kolmogorov spectrum in the inertial range, eddy diffusivity is predicted to scale as $\widehat{\mathcal{D}}\sim k^{-4/3}$ while our measurement over a range of Reynolds numbers consistently results in $\widehat{\mathcal{D}}\sim k^{-1}$. This is because the RNG-based model only considers the effects of eddies of size $1/k$ or smaller in mixing at wavenumber $k$. Our results suggest that large eddies contribute significantly to mixing at scales smaller than the eddy size.   This contrast can be partly understood by noting the Eulerian nature of RANS modeling framework: in an otherwise Lagrangian setting in which averaging is performed on a probe moving with the local flow, large eddies would dominantly advect the small scale structures and hardly mix them. But in our framework, to be relevant to RANS model expectations, time averaging is performed on a probe fixed in space. Therefore, advection by large eddies would lead to highly ``mixed" samplings of the transported quantities, hence a higher eddy diffusivity than those considered in RNG theories.
 
 
Our results also implicate closure models that use fractional order differential operators. The model presented in this study involves a fractional order operator due to the operational square root in the denominator of ~\ref{eq:D}. However, this model involves fractional order operations in a quantitatively distinct way compared to historic models in the literature. Majority of the studies in the literature consider fractional order closure operators in terms of a Laplacian operator raised to a non-integer power, i.e. represented by a single power law in Fourier space \cite{clark2021}. However, at least for HIT, our work shows that the closure operator is not a unified power law across all scales. We intuitively explained the shift in the power law from $D\sim k^0$ in the small $k$ limit to $D\sim k^{-1}$ in the large $k$ limit based on requirement of finite speed of advective transport at all scales. This requirement is not limited to HIT and thus standard single power law fractional-order operators are unlikely to be universal quantitative matches for practical turbulent flows.


\section{Acknowledgements}
The study presented in this report was supported by the Boeing company under grant number SPO 134136 an the U.S. Department of Energy, under grant number DE-NA0002373. 


\providecommand{\noopsort}[1]{}\providecommand{\singleletter}[1]{#1}%

\end{document}